\documentclass[conference]{IEEEtran}
\IEEEoverridecommandlockouts
\usepackage{cite}
\usepackage{amsmath,amssymb,amsfonts}
\usepackage{algorithmic}
\usepackage{graphicx}
\usepackage{textcomp}
\usepackage{xcolor}
\usepackage[hidelinks]{hyperref}
\usepackage{subcaption}
\usepackage{multirow}
\usepackage{balance}

\def\BibTeX{{\rm B\kern-.05em{\sc i\kern-.025em b}\kern-.08em
    T\kern-.1667em\lower.7ex\hbox{E}\kern-.125emX}}
\begin{document}

\title{CAT-LDP: Cloud-edge Adaptive Taxonomy under Local Differential Privacy\\
}

\author{\IEEEauthorblockN{1\textsuperscript{st} Junzhe Yang*}
\IEEEauthorblockA{\textit{Pittsburgh Institute} \\
\textit{Sichuan University}\\
Chengdu, China \\
and83870@gmail.com}
*Corresponding Author
~\\
\and
\IEEEauthorblockN{1\textsuperscript{st} Chang Xia}
\IEEEauthorblockA{\textit{Pittsburgh Institute} \\
\textit{Sichuan University}\\
Chengdu, China \\
xiachang64@gmail.com}

~\\

\and
\IEEEauthorblockN{2\textsuperscript{nd} Xiyun Wang}
\IEEEauthorblockA{\textit{Pittsburgh Institute} \\
\textit{Sichuan University}\\
Chengdu, China \\
2024141520230@stu.scu.edu.cn}

~\\

\and
\IEEEauthorblockN{3\textsuperscript{rd} Anren Sun}
\IEEEauthorblockA{\textit{Pittsburgh Institute} \\
\textit{Sichuan University}\\
Chengdu, China \\
2024141520234@stu.scu.edu.cn}

~\\

\and
\IEEEauthorblockN{4\textsuperscript{th} Wenbo Ding}
\IEEEauthorblockA{\textit{Pittsburgh Institute} \\
\textit{Sichuan University}\\
Chengdu, China \\
2024141520237@stu.scu.edu.cn}

~\\

\and
\IEEEauthorblockN{5\textsuperscript{th} Xinye Chen}
\IEEEauthorblockA{\textit{Pittsburgh Institute} \\
\textit{Sichuan University}\\
Chengdu, China \\
2024141520177@stu.scu.edu.cn}

}

\maketitle

\begin{abstract}
Recommender systems are widely used in daily life, but their direct collection and use of user preference data can also lead to privacy leakage. Existing privacy-preserving recommendation methods often find it hard to balance user privacy and recommendation performance. This problem is more serious in implicit-feedback settings, where data sparsity further increases the loss of useful signals caused by privacy perturbation. To solve this problem, we propose CAT-LDP, a cloud-local collaborative recommendation framework under local differential privacy constraints. CAT-LDP combines a hierarchical taxonomy tree with an adaptive privacy budget allocation strategy to keep more useful signals in users’ active categories while protecting user privacy. Specifically, users upload perturbed category profiles that satisfy LDP. Based on these profiles, the cloud performs coarse-grained candidate generation, and the local device then carries out fine-grained reranking by using unperturbed local history. Experiments on the Amazon Video Games dataset show that CAT-LDP consistently outperforms its fixed-budget ablation variant and representative baselines on HR@K and NDCG@K under different privacy budgets. The results show that combining category-space modeling with cloud-local task decoupling can effectively reduce noise amplification in long-tail sparse settings and provide a better balance between privacy and utility for implicit-feedback recommendation.
\end{abstract}


\section{Introduction}
Recommender systems are a representative form of personalized service and heavily rely on user behavioral data to drive recommendation algorithms. In particular, recommendation models estimate user preferences from collected behavioral data and generate candidate items for future consumption \cite{hu2008implicit,rendle2009bpr}. As a result, service providers operating recommender systems usually have full access to raw user data. Such data can reveal, or enable the inference of, sensitive user attributes. Therefore, recommender systems face significant risks of user privacy leakage \cite{calandrino2011privacy,xin2023behavior}.

To address this issue, privacy-preserving recommendation has been widely studied in recent years \cite{mullner2023dpreview,ali2025privacysurvey}. However, two challenges remain. First, protecting users against a curious or malicious server is still difficult in practical recommendation scenarios. Existing studies have shown that sensitive information may still be inferred from recommender-side information, including public outputs, exposure signals, and uploaded model components or updates \cite{calandrino2011privacy,xin2023behavior,zhang2024attribute}. Second, privacy-preserving mechanisms usually introduce a utility--privacy trade-off, because injected noise can weaken useful preference signals and reduce recommendation accuracy \cite{mullner2024impact}. This issue is more severe in implicit-feedback settings, where observations are inherently sparse and do not provide explicit negative preference labels \cite{hu2008implicit,rendle2009bpr}.

To address the above challenges, in this paper we propose a cloud-local decoupled collaborative tree-based recommendation framework for implicit-feedback scenarios. The proposed framework constructs a three-level hierarchical tree based on category information and assigns recommendation tasks of different granularities to the cloud and the local device, respectively. Each user uploads only privacy-protected upper-level data to the server under differential privacy, and only the third-level information is used to perform coarse-grained recommendation on the cloud side. This design reduces the privacy risk caused by treating the recommendation server as a potential adversary. Based on the coarse-grained results, the local device further performs fine-grained personalized recommendation using lower-level data, thereby alleviating the negative impact of data sparsity on recommendation accuracy while preserving user privacy.

\subsection{Related Work}
Traditional recommendation models primarily focus on improving recommendation utility in centralized environments. Collaborative filtering techniques, represented by matrix factorization, achieve high predictive performance by learning the latent vectors between users and items \cite{rendle2009bpr,salakhutdinov2007pmf,koren2009matrix}. Subsequently, to handle large-scale feature interactions and non-linear relationships, researchers introduced tree-structured recommendation models \cite{he2014practical,zhu2018learning}. These models capture complex interaction patterns through the hierarchical partitioning of the feature space. However, the aforementioned works are all built upon an implicit premise: the cloud server is fully trusted. This means that users' sensitive interaction information is completely exposed to the cloud server. In practice, this operation raises severe privacy concerns.

To protect users' sensitive interaction information from attacks by untrusted service providers, researchers introduced the concept of Local Differential Privacy (LDP) into traditional recommendation systems. Local differential privacy requires user data to be perturbed before leaving the local device \cite{kasiviswanathan2011can, erlingsson2014rappor}. A comprehensive review of LDP methods and their specific challenges in recommendation can be found in the recent survey by Yang et al. \cite{yang2024survey}. Targeting recommendation scenarios, Shin et al. proposed frameworks such as PrivPF \cite{shin2018privacy}, attempting to protect users' item sets and ratings under LDP constraints. Subsequently, works like LDPMF \cite{li2025ldpmf} further explored how to reconstruct matrix factorization models on the cloud after perturbing gradients or ratings locally, thereby further improving recommendation accuracy. More recently, to address the cascading accuracy degradation in traditional matrix factorization, Zhang et al. \cite{zhang2025privacy} proposed combining a bidirectional bounded LDP perturbation mechanism with a Gaussian-Laplacian mixture model. Tang et al. \cite{bayesian_ldp_2025} introduced a two-stage Bayesian LDP recommendation framework tailored for implicit feedback, which applies randomized response locally and leverages Bayesian inference on the server to effectively reconstruct interaction probabilities. 

To address the extremely low signal-to-noise ratio and the high computational cost of matrix operations in pure LDP scenarios, Gao et al. proposed the DPLCF model \cite{gao2020dplcf}. This marks a paradigm shift for LDP-based recommendation systems from "pure local perturbation" to "local perturbation uploading with cloud-local collaborative recommendation." DPLCF designed a novel general framework for differentially private collaborative filtering tailored to implicit feedback. It adopts widely used differential privacy standards during the data collection process, and on the cloud side, it infers item similarities solely from the protected data to serve as parameters for the item-based recommendation model. Finally, the item similarities are sent back to the device and combined with locally stored data to form an item-based collaborative filtering approach that generates the recommendation results.

Although these methods provide rigorous privacy guarantees and perform well in recommendation metrics, we still identified several limitations: 
\vspace{1.5mm}
\begin{itemize}
    \item \textbf{Vulnerability to long-tail datasets:} Existing LDP recommender methods typically perturb user interactions in a high-dimensional, flat item space. In highly sparse long-tail settings, this representation weakens effective signal density and increases noise sensitivity, making weak user-intent signals harder to recover. Without an explicit taxonomic structure, models have limited ability to aggregate collaborative evidence across semantically related items.
    \item \textbf{Fixed privacy budget allocation:} When allocating the privacy budget, the aforementioned models typically assign an identical privacy budget to all items. However, in real-world scenarios, different items or categories expose varying degrees of user privacy, and therefore should be allocated different privacy budgets.
    \item \textbf{Underutilization of Local Context:} Current cloud-local pipelines also use clean local context. Most methods perform final similarity estimation or scoring primarily on perturbed server-side data, rather than separating coarse retrieval and local reranking. As a result, unperturbed on-device histories, which provide reliable fine-grained preference cues, are not fully leveraged in the ranking stage.
    
\end{itemize}

\subsection{Contribution}
To address the aforementioned limitations, this paper proposes an adaptive cloud-local collaborative LDP recommendation model based on a category tree. On the local device, leveraging a stable category tree structure, the model first constructs user profiles using shallow categories to mitigate the curse of dimensionality inherent in extreme long-tail datasets. It then allocates an adaptive privacy budget, adds noise, and uploads the perturbed data. Upon receiving the noisy data, the server clusters the target users and generates coarse scores based on the clustering results. Subsequently, the coarsely ranked Top-M candidate items are dispatched back to the local device for fine-grained ranking, generating the final recommendation list. This approach utilizes a more personalized privacy protection method while effectively improving recommendation performance on extreme long-tail datasets.

The main contributions of this paper are summarized as follows:
\vspace{1.5mm}
\begin{itemize}
    \item We present CAT-LDP, a decoupled cloud-local recommendation framework for highly sparse implicit-feedback settings under LDP. The framework performs privacy-preserving coarse ranking on the cloud and metadata-assisted fine reranking on the local device, which improves the privacy and utility trade-off in this setting. 
    \item We introduce a hierarchical taxonomy tree representation to structure the feature space. We further design an adaptive privacy-budget allocation strategy that assigns larger budgets to user-active categories under a fixed mean-budget constraint. This design improves the quality of informative dimensions for downstream ranking.
    \item Experiments on the Amazon Video Games dataset show that CAT-LDP outperforms its fixed-budget ablation and representative LDP baselines on reported HR@K and NDCG@K metrics.
\end{itemize}

\section{PRELIMINARIES}

\subsection{Problem Formulation}

In our recommendation scenario, the user set is defined as $\mathcal{U}=\{u_1, u_2, \dots, u_{|\mathcal{U}|}\}$, and the item set as $\mathcal{I} = \{i_1, i_2, \dots, i_{|\mathcal{I}|}\}$. To alleviate the high-dimensional sparsity issue caused by extreme long-tail datasets, we introduce a Category Tree structure for items. Assuming the category tree consists of coarse-grained (Level-2) and fine-grained (Level-3) hierarchies, we denote the bottom-level fine-grained category set as $\mathcal{C} = \{c_1, c_2, \dots, c_{|\mathcal{C}|}\}$.

The true historical interaction profile of each user $u \in \mathcal{U}$ within the category space can be represented by a binary feature vector $\mathbf{x}_u = [x_{u,1}, x_{u,2}, \dots, x_{u,|\mathcal{C}|}] \in \{0, 1\}^{|\mathcal{C}|}$. Here, $x_{u,j} = 1$ indicates that the user has positive implicit feedback (e.g., clicking, playing) on category $c_j$, and $0$ otherwise.

\vspace{1.5mm}
\textbf{Privacy-Preserving Recommendation Goal:} 
Under the untrusted server setting, $\mathbf{x}_u$ is not directly uploaded. The system workflow operates as follows: the user's device applies local perturbation to $\mathbf{x}_u$ and uploads the noisy version $\tilde{\mathbf{x}}_u$; the cloud server performs clustering and coarse candidate ranking based exclusively on the collected noisy dataset $\{\tilde{\mathbf{x}}_u\}_{u \in \mathcal{U}}$; subsequently, the top $M$ candidates from the cloud's ranking are dispatched back to the local device for reranking. The local device then extracts public metadata tokens (e.g., title, brand, and category) of the candidate items and combines them with the unperturbed local historical records $\mathcal{H}_u$ to perform a lightweight reranking via Jaccard similarity computation. Our optimization objective is to enhance the final ranking quality (e.g., HR@$K$, NDCG@$K$) under privacy constraints.

\subsection{Local Differential Privacy}

Local Differential Privacy provides a mathematical guarantee for the local perturbation of user data. Its formal definition is as follows: 

\vspace{1.5mm}
\textbf{Definition:} 
Given a privacy budget $\epsilon \ge 0$, a randomized mechanism $\mathcal{M}$ satisfies $\epsilon$-LDP if and only if, for any two distinct input feature vectors $\mathbf{x}, \mathbf{x}'$ and any output $\tilde{\mathbf{x}}$, the following holds:
\begin{equation}
\Pr[\mathcal{M}(\mathbf{x}) = \tilde{\mathbf{x}}] \le e^\epsilon \Pr[\mathcal{M}(\mathbf{x}') = \tilde{\mathbf{x}}]
\end{equation}

For binary features, this paper adopts the Binary Randomized Response (BRR) mechanism:
\begin{equation}
\Pr[\tilde{x} = 1 \mid x = 1] = p = \frac{e^\epsilon}{1 + e^\epsilon}
\end{equation}
\begin{equation}
\Pr[\tilde{x} = 1 \mid x = 0] = q = \frac{1}{1 + e^\epsilon}
\end{equation}

In traditional fixed-budget paradigms, all dimensions share an identical baseline budget $\epsilon_{\text{fixed}}$. While straightforward to implement, when facing a high-dimensional and extremely sparse category space, this strategy often allocates a substantial portion of the budget to low-intent dimensions that are completely irrelevant to the user, leading to an underutilization of effective signals.

\subsection{Adaptive Budget Policy}
To improve the signal-to-noise ratio without altering the overall budget magnitude, we break the restriction of a uniform budget and allow the use of a dimensionalized budget vector $\boldsymbol{\epsilon}_u = [\epsilon_{u,1}, \dots, \epsilon_{u,|\mathcal{C}|}]$ for each user $u$.

To ensure that our adaptive mechanism maintains a comparable privacy overhead to the fixed-$\epsilon$ scheme under an average budget constraint, we impose two types of constraints on the generation of the budget vector:

\begin{itemize}
    \item \textbf{Boundary Constraint:} To prevent the over-exposure or over-perturbation of any single dimension, the budget allocation is restricted within specific upper and lower bounds. Specifically, the clipping boundaries scale proportionally with the global baseline budget $\epsilon_{\text{fixed}}$:
    \begin{equation}
    \epsilon_{\min} \le \epsilon_{u,j} \le \epsilon_{\max}, \quad \forall j \in \{1, \dots, |\mathcal{C}|\}
    \end{equation}
    where $\epsilon_{\min}$ and $\epsilon_{\max}$ are the scaling boundaries calculated based on $\epsilon_{\text{fixed}}$ (e.g., $\epsilon_{\min} = 0.5 \cdot \epsilon_{\text{fixed}}$, $\epsilon_{\max} = 4.0 \cdot \epsilon_{\text{fixed}}$).

    \item \textbf{Mean Equivalency Constraint:} For any given user $u$, the mean value of their adaptive budget vector must numerically approximate the global baseline budget:
    \begin{equation}
    \frac{1}{|\mathcal{C}|} \sum_{j=1}^{|\mathcal{C}|} \epsilon_{u,j} \approx \epsilon_{\text{fixed}}
    \end{equation}
\end{itemize}

\vspace{1.5mm}
\textbf{Implementation and Solution Process:} 
During the actual allocation, the model first constructs an initial hierarchical budget $\hat{\boldsymbol{\epsilon}}_u$ based on the user's activity in the Level-2 coarse categories (assigning higher budgets to highly active categories). Subsequently, to satisfy the mean equivalency constraint, the model seeks an optimal scaling factor $\alpha_u$ such that:
\begin{equation}
\frac{1}{|\mathcal{C}|} \sum_{j=1}^{|\mathcal{C}|} \text{clip}(\alpha_u \hat{\epsilon}_{u,j}, \epsilon_{\min}, \epsilon_{\max}) \approx \epsilon_{\text{fixed}}
\end{equation}
Because this objective function is monotonically increasing with respect to $\alpha_u$, the parameter $\alpha_u$ can be approximately solved within a predefined tolerance via binary search. This process essentially corresponds to ``budget reallocation under a fixed average overhead,'' effectively concentrating privacy resources on the user's high-intent features, thereby improving recommendation utility at a comparable privacy cost.

\section{System Architecture}
This paper proposes an adaptive cloud-local collaborative LDP recommendation framework based on a category tree. To strictly protect the privacy of users' implicit feedback while effectively alleviating the high-dimensional sparsity and severe noise interference caused by extreme long-tail datasets, the overall system architecture is designed as a "local-cloud-local" three-phase collaborative paradigm. The specific workflow can be divided into the following three core phases:

\begin{figure*}[t]
    \centering
    \includegraphics[width=0.8\textwidth]{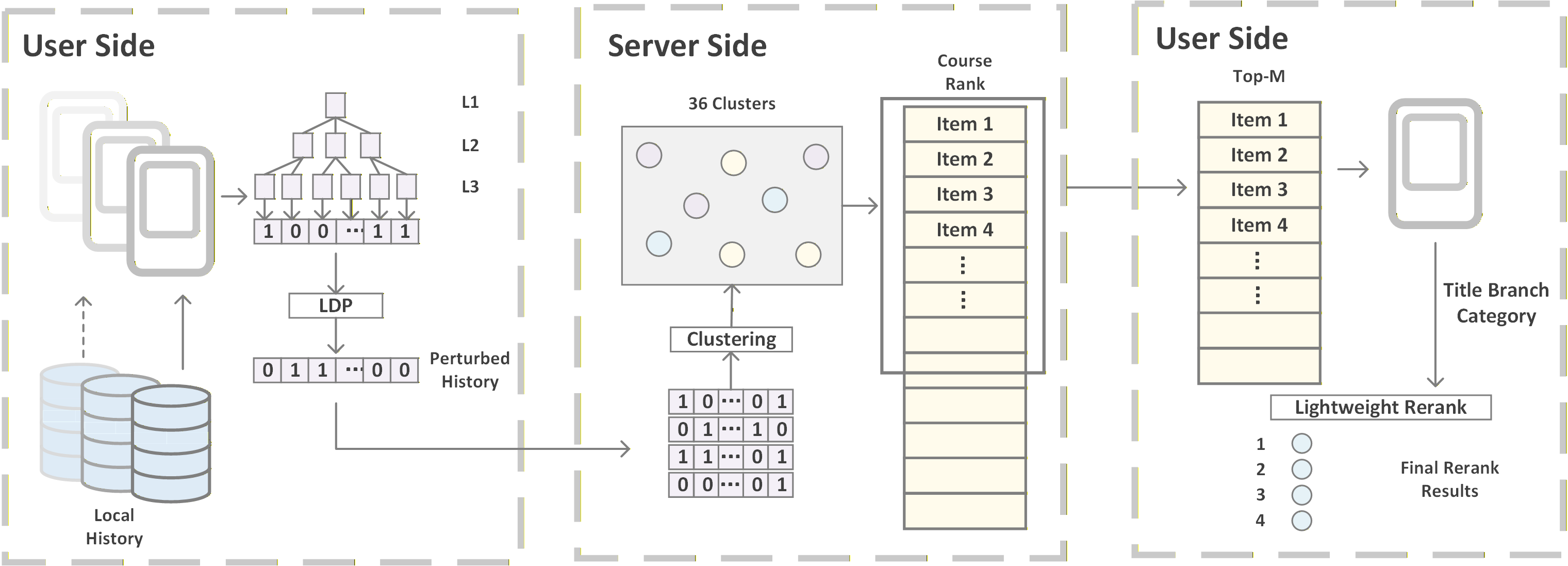}
    \caption{The Proposed Framework for CAT-LDP}
    \label{fig:framework}
\end{figure*}

\subsection{Local Profile Construction and Adaptive Perturbation}

On the user's local device, the system first maps their unperturbed, true historical interaction records $\mathcal{H}_u$ into the bottom-level fine-grained (Level-3) category space, constructing a high-dimensional binary feature profile $\mathbf{x}_u$. To enhance effective signal retention under a fixed average privacy overhead, the local module adopts a non-uniform perturbation strategy. Based on the user's activity in coarse-grained categories, it dynamically allocates a dimensionalized adaptive privacy budget $\boldsymbol{\epsilon}_u$ that satisfies the "mean equivalency constraint." Subsequently, the system perturbs $\mathbf{x}_u$ using the Binary Randomized Response (BRR) mechanism to generate an LDP-protected noisy vector $\tilde{\mathbf{x}}_u$, which is then uploaded to the untrusted cloud.\\
\textbf{Input:} User's historical records $\mathcal{H}_u$.\\
\textbf{Output:} LDP-protected noisy feature vector $\tilde{\mathbf{x}}_u$.

\subsection{Cloud-Side Collaborative Clustering and Coarse Ranking}

Upon receiving the noisy dataset $\{\tilde{\mathbf{x}}_u\}_{u \in \mathcal{U}}$ from all users, the cloud server faces the challenges of an extremely low signal-to-noise ratio and highly sparse data. To extract macroscopic collaborative filtering signals from the perturbed data, the server first clusters the target user group, leveraging group-level features to smooth out the random noise introduced by individual LDP. Subsequently, over a given set of candidate items, the cloud extracts the cluster center's feature representations corresponding to the items' categories and combines them with the user's category mask to execute coarse ranking scoring. Finally, the cloud dispatches the top $M$ candidate items with the highest coarse ranking scores to the target user's local device.\\
\textbf{Input:} Perturbed dataset from all users $\{\tilde{\mathbf{x}}_u\}_{u \in \mathcal{U}}$.\\
\textbf{Output:} Top-$M$ coarse candidate list.

\subsection{Local Lightweight Reranking}

Upon receiving the top $M$ candidate items dispatched by the cloud, the ranking process is handed back to the user's local device. The local module extracts the public metadata tokens (e.g., title, brand, and category) of these $M$ candidate items and directly retrieves the unperturbed, true interaction history $\mathcal{H}_u$ stored locally. By calculating the Jaccard similarity between the candidate items' metadata and the local historical records, and employing max aggregation, the local device performs a lightweight personalized reranking exclusively on these $M$ coarse candidates to output the final recommendation list.\\
\textbf{Input:} Top-$M$ candidate items.\\
\textbf{Output:} Final recommendation list.

\section{The proposed method}
This section details the technical implementation of the proposed recommendation framework called CAT-LDP. Based on the data flow, the entire model architecture is divided into three modules: local data preprocessing and perturbation, cloud-side collaborative clustering and coarse ranking, and local metadata-driven reranking.

\subsection{User Side 1}
Under the untrusted server setting, the model first performs feature space transformation and privacy perturbation locally on the user's device.

\vspace{1.5mm}
\subsubsection{User-Category Profile Construction}
Given the true local interaction history $\mathcal{H}_u$ of user $u$, the system extracts the fine-grained categories (Level-3) of the historical items to construct a high-dimensional binary feature vector $\mathbf{x}_u \in \{0, 1\}^{|\mathcal{C}|}$. If the user's history includes an item under category $c_j$, the corresponding dimension $x_{u,j} = 1$; otherwise, it is $0$.

\vspace{1.5mm}
\subsubsection{Adaptive Privacy Budget Allocation}
To alleviate budget waste on high-dimensional sparse features, the system allocates heterogeneous budgets to each dimension of $\mathbf{x}_u$ based on the "mean equivalency constraint." Let the set of coarse-grained Level-2 categories be $\mathcal{G}$, and the set of fine-grained Level-3 indices corresponding to the $g$-th Level-2 category be $\mathcal{I}_g$. The system first calculates the user's activity in each Level-2 category:
\begin{equation}
    a_{u,g} = \sum_{j \in \mathcal{I}_g} x_{u,j}, \quad g \in \mathcal{G}
\end{equation}
The top $L$ most active Level-2 categories are selected and denoted as $\mathcal{G}_u^{(L)}$. The system sets the initial hierarchical budgets according to activity levels (where $s_b, s_{bo}, s_d$ are the corresponding scaling coefficients):
\begin{equation}
    \begin{aligned}
    \epsilon_{\text{base}} &= \max(\epsilon_{\min}, \epsilon_{\text{fixed}} \cdot s_b), \\
    \epsilon_{\text{boost}} &= \min(\epsilon_{\max}, \epsilon_{\text{fixed}} \cdot s_{bo}), \\
    \epsilon_{\text{deep}} &= \min(\epsilon_{\max}, \max(\epsilon_{\text{boost}}, \epsilon_{\text{fixed}} \cdot s_d)).
    \end{aligned}
\end{equation}

For each dimension $j$, its unnormalized initial budget $\hat{\epsilon}_{u,j}$ is allocated based on its category hierarchy as follows:
\begin{equation}
    \hat{\epsilon}_{u,j} = \begin{cases}
    \epsilon_{\text{deep}}, & j \in \mathcal{D}_u \\
    \epsilon_{\text{boost}}, & j \in \bigcup_{g \in \mathcal{G}_u^{(L)}} \mathcal{I}_g, \, j \notin \mathcal{D}_u \\
    \epsilon_{\text{base}}, & \text{otherwise}
\end{cases}    
\end{equation}

\vspace{1.5mm}
\subsubsection{Symmetric Flipping Perturbation Logic}
After determining the budget $\epsilon_{u,j}$ for each dimension, the system applies a symmetric flipping mechanism based on randomized response to add bit-wise noise to $\mathbf{x}_u$. For the $j$-th dimension in the vector, the flipping probability parameters $p_{u,j}$ and $q_{u,j}$ are defined as:
\begin{equation}
    p_{u,j} = \frac{e^{\epsilon_{u,j}}}{1 + e^{\epsilon_{u,j}}}, \quad q_{u,j} = \frac{1}{1 + e^{\epsilon_{u,j}}}
\end{equation}

The system samples a random number $r \sim U(0,1)$ from a uniform distribution and generates the noisy indicator $\tilde{x}_{u,j}$ to be uploaded to the cloud according to the following piecewise function:
\begin{equation}
    \tilde{x}_{u,j} = \begin{cases}
    1, & \text{if } r \le p_{u,j} \text{ and } x_{u,j} = 1 \\
    0, & \text{if } r > p_{u,j} \text{ and } x_{u,j} = 1 \\
    0, & \text{if } r > q_{u,j} \text{ and } x_{u,j} = 0 \\
    1, & \text{if } r \le q_{u,j} \text{ and } x_{u,j} = 0
    \end{cases}
\end{equation}

This mechanism ensures that the flipping process satisfies the LDP definition. The output noisy vector $\tilde{\mathbf{x}}_u$ is then transmitted to the cloud server.

where $\mathcal{D}_u$ represents the top $m$ active Level-3 dimensions selected within each triggered Level-2 category. Subsequently, an optimal scaling factor $\alpha_u$ is determined via binary search to perform mean equivalency normalization on the budget. This ensures that the clipped budget for each dimension, $\epsilon_{u,j} = \text{clip}(\alpha_u \hat{\epsilon}_{u,j}, \epsilon_{\min}, \epsilon_{\max})$, satisfies:
\begin{equation}
    \frac{1}{|\mathcal{C}|} \sum_{j=1}^{|\mathcal{C}|} \epsilon_{u,j} \approx \epsilon_{\text{fixed}}
\end{equation}

\subsection{Server Side}
Upon receiving the noisy matrix $X' = \{\tilde{\mathbf{x}}_1, \dots, \tilde{\mathbf{x}}_{|\mathcal{U}|}\}$ from all users, the cloud server's core task is to extract collaborative filtering signals in an extremely low signal-to-noise ratio environment and perform an initial screening of the candidate pool.

\vspace{1.5mm}
\subsubsection{Robust Graph-Based Consensus Clustering}
Calculating user similarity directly based on $\tilde{\mathbf{x}}_u$ leads to severe ranking bias. Therefore, the cloud adopts a multi-stage ensemble clustering strategy. The system first independently executes multiple sets of K-Means clustering with different cluster number configurations (e.g., $k \in \{30, 36, 42\}$) on $X'$, obtaining a deduplicated set of base clusters $\{\mathcal{S}_b\}_{b=1}^B$. For each base cluster $\mathcal{S}_b$, its cluster center is calculated as $\boldsymbol{\mu}_b = \frac{1}{|\mathcal{S}_b|} \sum_{u \in \mathcal{S}_b} \tilde{\mathbf{x}}_u$.

Subsequently, the system constructs an inter-cluster graph by integrating the micro-cluster center similarity $C_{bb'}$ and the member overlap relationship $O_{bb'}$. The relevant metrics and the size regularization term $\beta_b$ are defined as follows:
\begin{equation}
    C_{bb'} = \max(\cos(\boldsymbol{\mu}_b, \boldsymbol{\mu}_{b'}), 0)
\end{equation}

\begin{equation}
    O_{bb'} = \frac{|\mathcal{S}_b \cap \mathcal{S}_{b'}|}{|\mathcal{S}_b \cup \mathcal{S}_{b'}|}
\end{equation}

\begin{equation}
    \beta_b = \frac{1}{1 + \log(\max(|\mathcal{S}_b|, 1))}
\end{equation}

This yields the inter-cluster graph weight matrix:
\begin{equation}
    W_{bb'} = \beta_b \beta_{b'} (0.7 C_{bb'} + 0.3 O_{bb'})
\end{equation}

Spectral Clustering is then performed on $W$ to obtain the mapping $g(b) = c$ from base clusters to final clusters. The member set of the final cluster $c$ and its corresponding center vector $\mathbf{v}_c$, which reduces the variance caused by random perturbation, are respectively denoted as:
\begin{equation}
    \mathcal{F}_c = \bigcup_{b: g(b)=c} \mathcal{S}_b, \quad \mathbf{v}_c = \frac{1}{|\mathcal{F}_c|} \sum_{u \in \mathcal{F}_c} \tilde{\mathbf{x}}_u
\end{equation}

After clustering, user $u$ is assigned to the final cluster with the highest cosine similarity:
\begin{equation}
    C_u = \arg\max_c \cos(\tilde{\mathbf{x}}_u, \mathbf{v}_c)
\end{equation}

\vspace{1.5mm}
\subsubsection{Multiplicative Gated Coarse Ranking}
When generating the candidate list for user $u$, the cloud performs coarse ranking scoring for candidate item $i$. Let the category of item $i$ be $c_i$, and the user's category value visible to the cloud be $z_{u, c_i}$ (corresponding to $\tilde{x}_{u, c_i}$ under the LDP branch). To combine global collaborative signals with individual intent, the system employs the following multiplicative gated scoring mechanism:
\begin{equation}
    S_{\text{coarse}}(u, i) = \max(v_{C_u, c_i}, 0) \cdot (\lambda + (1 - \lambda) z_{u, c_i})
\end{equation}

Here, $\max(\cdot, 0)$ filters out negative preferences, and $\lambda \in [0, 1]$ is used to proportionally down-weight the cluster preference score for items lacking a positive indicator.

\subsection{User Side 2}
Upon receiving the candidate set dispatched by the cloud, the local device performs a secondary ranking utilizing the true historical records stored on the device.

\vspace{1.5mm}
\subsubsection{Metadata Feature Extraction}
For candidate item $i$ and historical interaction item $h \in \mathcal{H}_u$, the system extracts their public text metadata (title, brand, category, etc.) and maps them into token sets, denoted as $T(i)$ and $T(h)$ respectively.

\vspace{1.5mm}
\subsubsection{Max-Similarity Aggregation Scoring and Joint Ranking}
The system calculates the Jaccard similarity between candidate item $i$ and each historical item, employing a max aggregation strategy to extract the strongest matching signal as the local reranking score $S_{\text{local}}(u, i)$:
\begin{equation}
    S_{\text{local}}(u, i) = \max_{h \in \mathcal{H}_u} \frac{|T(i) \cap T(h)|}{|T(i) \cup T(h)|}
\end{equation}

Finally, the local device only reranks the top $M$ candidate items dispatched from the coarse ranking. The sorting key is set as $(-S_{\text{local}}, -S_{\text{coarse}})$, meaning items are primarily sorted in descending order based on their local score, referencing the coarse ranking score in the event of a tie. The tail un-reranked items maintain their cloud-side coarse ranking order. The overall list is then concatenated, and the Top-$K$ items are truncated to serve as the final output recommendation list.

\subsection{Complexity Analysis}
To demonstrate the efficiency and deployment feasibility of our proposed cloud-local collaborative framework, we analyze the theoretical time and space complexity across the three main processing phases.

\vspace{1.5mm}
\subsubsection{User Side 1}
\begin{itemize}
    \item \textbf{Time Complexity:} 
    Extracting the Level-3 profile and calculating Level-2 activity takes $\mathcal{O}(|\mathcal{C}| + |\mathcal{G}| \log |\mathcal{G}|)$. Since the number of coarse categories is much smaller than fine-grained ones ($|\mathcal{G}| \ll |\mathcal{C}|$), the dominant term remains $\mathcal{O}(|\mathcal{C}|)$. The budget allocation (via binary search) and the BRR perturbation execute in $\mathcal{O}(|\mathcal{C}|)$, leading to an overall time complexity dominated by $\mathcal{O}(|\mathcal{C}|)$.\\
    \item \textbf{Space Complexity:} 
    The local device only needs to store the sparse binary vector $\mathbf{x}_u$, the budget vector $\boldsymbol{\epsilon}_u$, and the perturbed vector $\tilde{\mathbf{x}}_u$. The space complexity is $\mathcal{O}(|\mathcal{C}|)$, which requires low memory overhead for modern mobile devices.
\end{itemize}

\vspace{1.5mm}
\subsubsection{Server Side}
\begin{itemize}
    \item \textbf{Time Complexity:} 
    The server-side computation is decoupled into offline clustering and online serving. For the \textbf{offline clustering}, executing K-Means across a set of cluster configurations $\mathcal{K}$ with $n_{\text{init}}$ initializations and $I_{\text{km}}$ iterations takes $\mathcal{O}(n_{\text{init}} I_{\text{km}} |\mathcal{U}| |\mathcal{C}| \sum_{k \in \mathcal{K}} k)$. Constructing the inter-cluster graph involves computing center similarities and member overlaps among the $B_0$ deduplicated base clusters, taking $\mathcal{O}(B_0^2 |\mathcal{C}| + B_0^2 |\mathcal{U}|)$. Performing Spectral Clustering takes $\mathcal{O}(B_0^3)$. Crucially, these operations are executed offline periodically. For \textbf{online coarse ranking}, computing the multiplicative gated scores and sorting the retrieval candidate set $\mathcal{I}_{\text{cand}}$ requires $\mathcal{O}(|\mathcal{I}_{\text{cand}}| \log |\mathcal{I}_{\text{cand}}|)$, which satisfies online response requirements.\\
    \item \textbf{Space Complexity:} 
    The server stores the noisy matrix from all users, the final cluster centers, and the inter-cluster similarity matrix $W$, requiring $\mathcal{O}(|\mathcal{U}| |\mathcal{C}| + K_{\text{final}} |\mathcal{C}| + B_0^2)$ storage space.
\end{itemize}

\vspace{1.5mm}
\subsubsection{User Side 2}
\begin{itemize}
    \item \textbf{Time Complexity:} 
    The local reranking phase computes the max Jaccard similarity between the top $M$ coarse candidates and the user's unperturbed history $\mathcal{H}_u$. Assuming the average number of metadata tokens per item is $|T|$, the time complexity is $\mathcal{O}(M \cdot |\mathcal{H}_u| \cdot |T|)$. Under scenarios where the average history length is controlled, this overhead can be approximated as lightweight.\\
    \item \textbf{Space Complexity:} 
    The device temporarily caches the metadata tokens of the $M$ downloaded candidates, taking $\mathcal{O}(M \cdot |T|)$ space. Once the ranking is complete, this cache can be immediately freed.
\end{itemize}

\section{Experiment}

\subsection{Experimental Settings}

\vspace{1.5mm}
\subsubsection{Datasets}
The experiments are conducted on the widely used public Amazon Product Data\cite{ni2019just}, specifically utilizing the \textbf{Video Games (5-core)} subset and its corresponding metadata. The Amazon dataset contains implicit feedback records from real users along with rich item attribute information.

In the context of our proposed framework, this dataset is adopted to construct the cloud-local collaborative recommendation scenario:

\begin{itemize}
    \item \textbf{Interaction Data:} The 5-core subset contains the historical records of active users (with at least 5 interactions). To align with our implicit feedback modeling setup, we binarize the explicit ratings in the original dataset into implicit interaction records, which constitute the unperturbed local historical sequence $\mathcal{H}_u$.
    \item \textbf{Category Tree and Metadata:} We leverage the hierarchical category information from the metadata to construct the fine-grained binary feature profile $\mathbf{x}_u$. Meanwhile, public text tokens such as titles and brands are extracted to support the Jaccard similarity computation during the local reranking phase.
\end{itemize}

To ensure a rigorous evaluation, we split the dataset into training, validation, and test sets. Specifically, we adopt the widely used \textit{leave-two-out} evaluation protocol: based on the chronological interaction sequence of each user, the most recent interaction is held out for testing, the second most recent for validation, and all remaining historical interactions are kept for training. Table \ref{tab:dataset_overall} presents the overall scale and the detailed statistics of the training set, while Table \ref{tab:dataset_split} reports the scale of the validation and test splits. As shown in Table \ref{tab:dataset_overall}, the sparsity of the training set exceeds 99.96\%, which reflects the extreme long-tail and high-dimensional sparsity challenges encountered in real-world recommender systems.

\begin{table}[t]
    \centering
    \caption{Statistics of the Video Games dataset}
    \label{tab:dataset_overall}
    \begin{tabular}{|c|c|c|c|c|}
    \hline
    \textbf{Dataset Split} & \textbf{\#Users} & \textbf{\#Items} & \textbf{\#Interactions} & \textbf{Sparsity} \\ \hline\hline
    Overall & 54,980 & 16,411 & 463,640 & 99.9486\% \\ \hline
    Training Set & 54,980 & 16,411 & 354,013 & 99.9608\% \\ \hline
    \end{tabular}
\end{table}

\begin{table}[t]
    \centering
    \caption{Scale of the Validation and Test Splits}
    \label{tab:dataset_split}
    \begin{tabular}{|c|c|c|c|}
    \hline
    \textbf{Split} & \textbf{\#Users} & \textbf{\#Items} & \textbf{\#Interactions} \\ \hline\hline
    Validation & 54,909 & 12,499 & 54,909 \\ \hline
    Test & 54,718 & 12,485 & 54,718 \\ \hline
    \end{tabular}
\end{table}

\vspace{1.5mm}
\subsubsection{Evaluation Metrics}

To objectively evaluate the recommendation performance of the models, we select two standard metrics widely adopted in the recommender system domain: Hit Ratio (HR@K) and Normalized Discounted Cumulative Gain (NDCG@K).

Under the data splitting protocol adopted in this paper, only one true interacted item (ground truth) is held out for each user in the test set $\mathcal{U}_{\text{test}}$. Based on this setting, and adopting a 1-based ranking index (i.e., $r_u \in \{1, 2, \dots\}$), the specific calculation logic for the two metrics is defined as follows: 

\begin{itemize}
    \item \textbf{HR@K:} 
    It is utilized to measure whether the recommender system successfully retrieves the target item into the top-$K$ recommendation list for a user. For a given user $u$, if the true item from their test set appears in the generated top-$K$ list, the hit score is 1; otherwise, it is 0. The overall HR@K of the system is the arithmetic mean of the scores across all test users:
    \begin{equation}
    \text{HR@K} = \frac{1}{|\mathcal{U}_{\text{test}}|} \sum_{u \in \mathcal{U}_{\text{test}}} \mathbf{1}[r_u \le K]
    \end{equation}
    where $r_u$ denotes the ranking position of the true target item for user $u$ in the recommendation list, and $\mathbf{1}[\cdot]$ is the indicator function, which equals 1 if the condition inside the brackets holds true, and 0 otherwise.

    \item \textbf{NDCG@K:} 
    Compared to HR, which only focuses on whether an item is hit, NDCG@K further examines the specific position of the hit item within the recommendation list. NDCG assigns higher score weights to hits ranked at the top based on a position decay logic. Since there is only one target item for each test user, the Ideal Discounted Cumulative Gain (IDCG) is always 1. Therefore, if the target item appears in the top-$K$ list, the user's score is the position-decayed value; if there is no hit, the score is 0. The formula is calculated as:
    \begin{equation}
    \text{NDCG@K} = \frac{1}{|\mathcal{U}_{\text{test}}|} \sum_{u \in \mathcal{U}_{\text{test}}} \frac{\mathbf{1}[r_u \le K]}{\log_2(r_u + 1)}
    \end{equation}
\end{itemize}

\vspace{1.5mm}
\subsubsection{Baselines}

Before introducing the comparative models, we first clarify the construction of the candidate set during the testing phase. We adopt the 1+99 sampled evaluation protocol, which is commonly used in recommender systems\cite{he2017neural}. Specifically, for each test user, we pair their single ground-truth interacted item (1 positive sample) with 99 un-interacted items (99 negative samples) randomly sampled from the training set to construct a test candidate set of size 100. To ensure comparability in the evaluation, all baseline methods and our proposed approach solely score and rank items within this specific candidate set. The HR@K and NDCG@K metrics are calculated based on the ranking of the target item within this candidate set.

Under this unified setting, to evaluate the effectiveness of the proposed method under privacy constraints and highly sparse scenarios, we select the following representative methods as baselines:

\begin{itemize}
    \item \textbf{CT-LDP:} 
    An ablation variant of the proposed framework. It retains the category tree profile, cloud-side clustering for coarse ranking, and local reranking processes, merely replacing the adaptive budget with fixed budget perturbation to evaluate the gain of the adaptive budget mechanism.
    \item \textbf{LCF-SP:} 
    An LCF baseline with symmetric perturbation ($p=1-q$). The cloud side does not perform unbiased noise correction and directly computes similarities based on the perturbed data\cite{gao2020dplcf}.
    \item \textbf{LCF-AP:} 
    An LCF baseline with asymmetric perturbation. Similarly, the cloud side does not perform unbiased correction and directly computes similarities based on the perturbed data\cite{gao2020dplcf}.
    \item \textbf{DPLCF-SP:} 
    The DPLCF method with symmetric perturbation. It introduces unbiased estimation for symmetric noise on the cloud side, and then combines it with the local history to generate recommendations\cite{gao2020dplcf}.
    \item \textbf{DPLCF-AP:} 
    The asymmetric variant of DPLCF. It adopts asymmetric perturbation locally, and uses the corresponding inverse estimation on the cloud side for bias correction to complete the recommendation\cite{gao2020dplcf}.
\end{itemize}

\vspace{1.5mm}
\subsubsection{Parameter settings}
To ensure the fairness of the experiments and the reproducibility of the results, this section details the core hyperparameter settings for the proposed method and the comparative baselines. For all algorithms, the processes involving random sampling are uniformly fixed with a random seed to eliminate performance fluctuations.

\begin{itemize}
    \item \textbf{Privacy perturbation parameters} 
    The base privacy budget of the framework is set to $\epsilon = 0.7$. In the adaptive budget allocation module, to avoid extreme flipping probabilities, the upper and lower bounds of the budget for single-dimensional features are strictly truncated, set to 0.5 times (i.e., $\epsilon_{\min} = 0.35$) and 4.0 times (i.e., $\epsilon_{\max} = 2.8$) the base budget, respectively.
    \item \textbf{Cloud collaboration parameters:} 
    In the offline graph clustering stage, the set of cluster numbers for K-Means base clustering is set to $\{30, 36, 42\}$, and the number of global clusters finally output by spectral clustering is set to $K_{\text{final}} = 36$. For the online coarse ranking computation, the penalty decay weight in the multiplicative gating function is set to $\lambda = 0.3$.
    \item \textbf{Local reranking parameters:} 
    The truncation depth of the local reranking candidate set is set to 20. The metadata similarity computation adopts the Jaccard index, and uses the maximum value to aggregate historical token scores.
    \item \textbf{Baseline Algorithm Parameters:} 
    For local collaborative filtering baselines such as LCF-SP, LCF-AP, DPLCF-SP, and DPLCF-AP, the item neighborhood size, which determines the scoring range and computational overhead, is uniformly set to $N = 20$, with a similarity block size of 256 and a perturbation batch size of 1024.
    
    To ensure fairness in the comparison, all baseline evaluations applicable to the hybrid local rerank mechanism adopt the same reranking depth = 20 and Jaccard similarity computation strategy as the proposed method.
\end{itemize}

\subsection{Performance Comparison}

\begin{figure*}[htbp]
    \centering
    \includegraphics[width=0.85\textwidth]{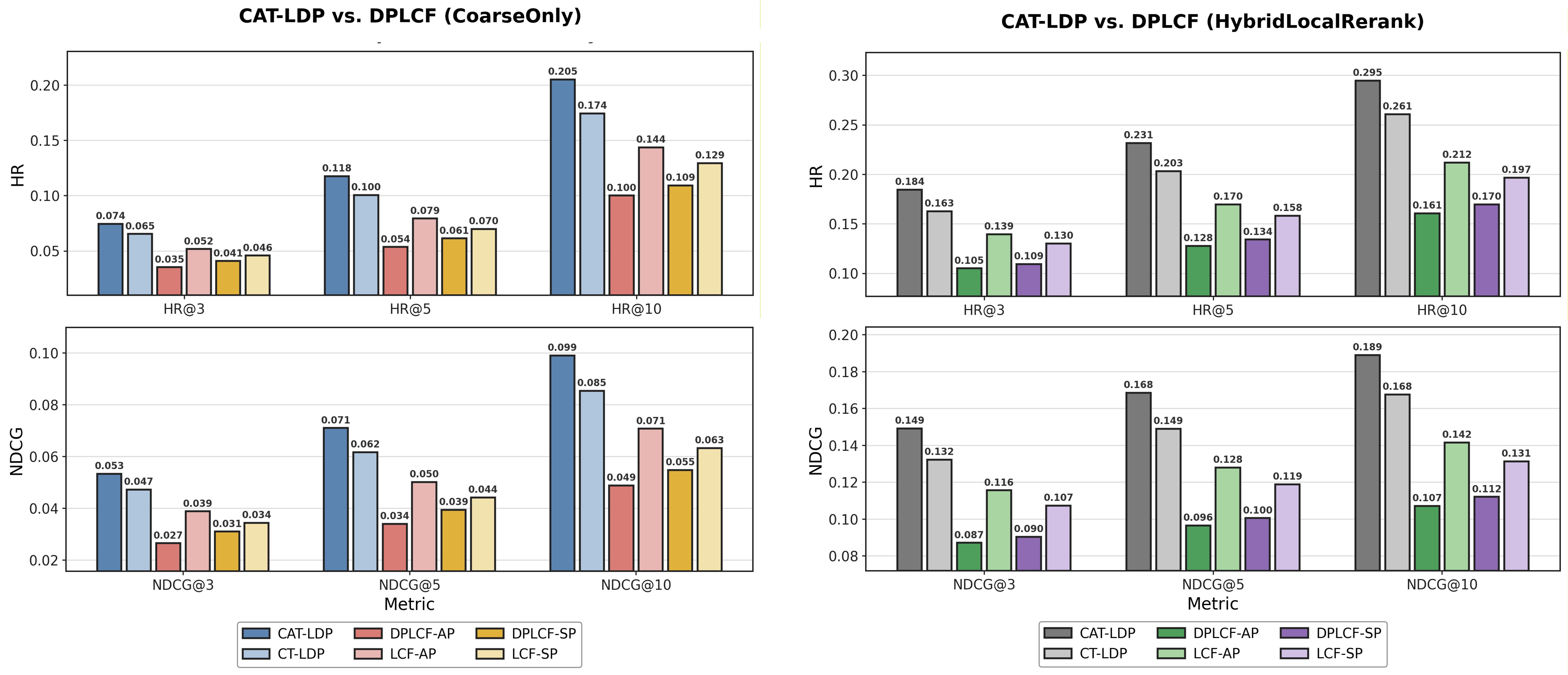}
    \caption{HR / NDCG performance comparison on the Amazon Video Game 5-core dataset.}
    \label{fig:Ours_vs_Baseline}
\end{figure*}

This section presents the performance evaluation results of our proposed CAT-LDP model, its ablation variant, and various baselines under the privacy budget $\epsilon=1.0$, considering both CoarseOnly and HybridLocalRerank scenarios, as shown in Figure 2 and Table 3 to 6. Overall, the proposed method achieves the best performance across all reported metrics ($K=2 \sim 10$) at different stages under the current experimental setting ($\epsilon=1.0$ and the Video Games dataset).

\vspace{1.5mm}
\subsubsection{Performance Analysis in the CoarseOnly Scenario}

In the setting that solely relies on the coarse ranking results returned from the cloud, CAT-LDP consistently performs the best across all truncation lengths $K$. For instance, at $K=10$, the HR@10 and NDCG@10 of CAT-LDP reach 0.2050 and 0.0990, respectively. 

Compared to the ablation variant CT-LDP which uses uniform budget allocation, CAT-LDP achieves performance improvements of approximately 17.5\% in HR@10 and 16.0\% in NDCG@10.

Compared to the best-performing baseline LCF-AP, CAT-LDP improves HR@10 by approximately 42.7\% and NDCG@10 by approximately 40.0\%. The improvement is equally significant when compared to the DPLCF series with unbiased correction.

\vspace{1.5mm}
\subsubsection{Performance Analysis in the HybridLocalRerank Scenario}

After introducing the metadata-based local reranking stage, the recommendation metrics for all algorithms exhibit substantial growth. This verifies the effectiveness of utilizing the unperturbed local history for Jaccard similarity reranking on this dataset.

Compared to the ablation variant CT-LDP, CAT-LDP maintains a steady performance advantage of approximately 13.0\% and 12.7\% in HR@10 and NDCG@10, respectively. 

Compared to the best baseline LCF-AP, our model yields scores that are approximately 39.1\% and 33.5\% higher, respectively.

\vspace{1.5mm}
\subsubsection{Analysis of Performance Improvements}

The advantages of our model over the baselines primarily stem from two aspects:

\begin{itemize}
    \item \textbf{Gains from adaptive budget allocation:} In the experiments, CAT-LDP consistently outperforms CT-LDP, demonstrating the effectiveness of dynamic budget allocation. In long-tail recommendation scenarios, user interests are typically concentrated in a few categories. By allocating more privacy budget to users' active coarse-grained categories and high-frequency fine-grained categories, CAT-LDP reduces budget consumption on irrelevant features. Consequently, under the same total privacy cost, it improves the signal-to-noise ratio of core feature vectors, helping provide a more reliable data foundation for accurate clustering on the cloud.
    \item \textbf{Advantages of category tree dimensionality reduction and cloud-local collaboration:} The DPLCF baselines inject randomized response noise directly at the item level. On sparse datasets, this may increase false positive records, degrading the quality of the item co-occurrence matrix. In contrast, our model first maps items to a hierarchical category tree and adds noise after feature dimensionality reduction, which helps alleviate the vulnerability to noise caused by high-dimensional sparsity. Meanwhile, compared to local item-item similarity computation, the global clustering mechanism on the cloud can capture group collaborative signals more stably.
\end{itemize}

\vspace{1.5mm}
\subsubsection{Analysis of Performance Degradation Caused by Bias Correction}

An observation from Tables 3 to 6 is that the LCF variants without unbiased estimation correction consistently outperform the DPLCF variants with unbiased correction. This result can be explained by the \textit{bias-variance trade-off} in high-dimensional sparse scenarios.

The core of DPLCF is to utilize flipping probabilities to construct the inverse of the state transition matrix, thereby obtaining an unbiased estimate of the true interaction counts. While this mechanism statistically guarantees the correctness of the expectation, its inverse algebraic operations amplify the variance of the estimates. In the dataset used in this paper, the true co-occurrence frequency between most item pairs is extremely low. In this case, the random noise amplified by the inverse matrix can easily overshadow the weak true signals. Although the computed similarities are strictly clipped to the $[0, 1]$ range in the code implementation, such drastic fluctuations in the underlying values still lead to a decrease in Top-$K$ ranking stability. Conversely, LCF directly computes the Jaccard similarity using the perturbed data. Although this approach introduces systemic bias, it acts as an implicit smoothing/regularization mechanism that controls variance inflation. In Top-$K$ retrieval tasks where evaluation metrics only focus on relative positions, the biased scores with stable relative magnitudes exhibit better ranking robustness under our experimental settings.

\begin{table*}[htbp]
\centering
\small
\caption{HR performance comparison on Amazon Video Game ($\epsilon=1.0$, CoarseOnly).}
\label{tab:coarseonly_hr_eps1}
\resizebox{0.92\textwidth}{!}{%
\begin{tabular}{l|c|c|c|c|c|c|c|c|c}
\hline
TopK & 2 & 3 & 4 & 5 & 6 & 7 & 8 & 9 & 10 \\
\hline
CT-LDP   & 0.0459 & 0.0653 & 0.0839 & 0.1004 & 0.1160 & 0.1317 & 0.1464 & 0.1604 & 0.1744 \\
CAT-LDP  & \textbf{0.0510} & \textbf{0.0744} & \textbf{0.0970} & \textbf{0.1175} & \textbf{0.1360} & \textbf{0.1542} & \textbf{0.1713} & \textbf{0.1883} & \textbf{0.2050} \\
LCF-AP   & 0.0372 & 0.0516 & 0.0663 & 0.0792 & 0.0927 & 0.1063 & 0.1189 & 0.1319 & 0.1436 \\
DPLCF-AP & 0.0257 & 0.0352 & 0.0442 & 0.0535 & 0.0630 & 0.0725 & 0.0814 & 0.0904 & 0.1000 \\
LCF-SP   & 0.0332 & 0.0457 & 0.0580 & 0.0697 & 0.0819 & 0.0943 & 0.1066 & 0.1176 & 0.1294 \\
DPLCF-SP & 0.0301 & 0.0407 & 0.0511 & 0.0612 & 0.0709 & 0.0804 & 0.0901 & 0.0996 & 0.1093 \\
\hline
\end{tabular}%
}
\end{table*}

\begin{table*}[htbp]
\centering
\small
\caption{NDCG performance comparison on Amazon Video Game ($\epsilon=1.0$, CoarseOnly).}
\label{tab:coarseonly_ndcg_eps1}
\resizebox{0.92\textwidth}{!}{%
\begin{tabular}{l|c|c|c|c|c|c|c|c|c}
\hline
TopK & 2 & 3 & 4 & 5 & 6 & 7 & 8 & 9 & 10 \\
\hline
CT-LDP   & 0.0375 & 0.0472 & 0.0552 & 0.0616 & 0.0672 & 0.0724 & 0.0770 & 0.0813 & 0.0853 \\
CAT-LDP  & \textbf{0.0416} & \textbf{0.0533} & \textbf{0.0630} & \textbf{0.0710} & \textbf{0.0776} & \textbf{0.0836} & \textbf{0.0890} & \textbf{0.0941} & \textbf{0.0990} \\
LCF-AP   & 0.0316 & 0.0388 & 0.0451 & 0.0501 & 0.0549 & 0.0595 & 0.0635 & 0.0674 & 0.0707 \\
DPLCF-AP & 0.0217 & 0.0265 & 0.0304 & 0.0340 & 0.0374 & 0.0405 & 0.0433 & 0.0461 & 0.0488 \\
LCF-SP   & 0.0281 & 0.0344 & 0.0396 & 0.0442 & 0.0485 & 0.0526 & 0.0565 & 0.0599 & 0.0633 \\
DPLCF-SP & 0.0257 & 0.0310 & 0.0355 & 0.0394 & 0.0429 & 0.0460 & 0.0491 & 0.0519 & 0.0547 \\
\hline
\end{tabular}%
}
\end{table*}

\begin{table*}[htbp]
\centering
\small
\caption{HR performance comparison on Amazon Video Game ($\epsilon=1.0$, HybridLocalRerank).}
\label{tab:hybridlocalrerank_hr_eps1}
\resizebox{0.92\textwidth}{!}{%
\begin{tabular}{l|c|c|c|c|c|c|c|c|c}
\hline
TopK & 2 & 3 & 4 & 5 & 6 & 7 & 8 & 9 & 10 \\
\hline
CT-LDP   & 0.1334 & 0.1625 & 0.1849 & 0.2031 & 0.2187 & 0.2315 & 0.2425 & 0.2518 & 0.2605 \\
CAT-LDP  & \textbf{0.1493} & \textbf{0.1845} & \textbf{0.2105} & \textbf{0.2315} & \textbf{0.2486} & \textbf{0.2629} & \textbf{0.2747} & \textbf{0.2849} & \textbf{0.2946} \\
LCF-AP   & 0.1165 & 0.1395 & 0.1565 & 0.1695 & 0.1807 & 0.1891 & 0.1977 & 0.2055 & 0.2117 \\
DPLCF-AP & 0.0880 & 0.1049 & 0.1176 & 0.1276 & 0.1367 & 0.1436 & 0.1496 & 0.1552 & 0.1605 \\
LCF-SP   & 0.1085 & 0.1301 & 0.1456 & 0.1580 & 0.1688 & 0.1770 & 0.1843 & 0.1906 & 0.1966 \\
DPLCF-SP & 0.0912 & 0.1093 & 0.1229 & 0.1340 & 0.1433 & 0.1510 & 0.1582 & 0.1641 & 0.1695 \\
\hline
\end{tabular}%
}
\end{table*}

\begin{table*}[htbp]
\centering
\small
\caption{NDCG performance comparison on Amazon Video Game ($\epsilon=1.0$, HybridLocalRerank).}
\label{tab:hybridlocalrerank_ndcg_eps1}
\resizebox{0.92\textwidth}{!}{%
\begin{tabular}{l|c|c|c|c|c|c|c|c|c}
\hline
TopK & 2 & 3 & 4 & 5 & 6 & 7 & 8 & 9 & 10 \\
\hline
CT-LDP   & 0.1176 & 0.1322 & 0.1419 & 0.1489 & 0.1545 & 0.1587 & 0.1622 & 0.1650 & 0.1675 \\
CAT-LDP  & \textbf{0.1315} & \textbf{0.1491} & \textbf{0.1603} & \textbf{0.1684} & \textbf{0.1745} & \textbf{0.1793} & \textbf{0.1830} & \textbf{0.1861} & \textbf{0.1889} \\
LCF-AP   & 0.1040 & 0.1155 & 0.1229 & 0.1279 & 0.1319 & 0.1347 & 0.1374 & 0.1397 & 0.1415 \\
DPLCF-AP & 0.0786 & 0.0871 & 0.0925 & 0.0964 & 0.0996 & 0.1019 & 0.1038 & 0.1055 & 0.1071 \\
LCF-SP   & 0.0964 & 0.1073 & 0.1139 & 0.1187 & 0.1226 & 0.1253 & 0.1276 & 0.1295 & 0.1312 \\
DPLCF-SP & 0.0813 & 0.0903 & 0.0962 & 0.1005 & 0.1038 & 0.1064 & 0.1087 & 0.1104 & 0.1120 \\
\hline
\end{tabular}%
}
\end{table*}

\subsection{Ablation Study}

\begin{figure*}[htbp]
    \centering
    \includegraphics[width=0.85\textwidth]{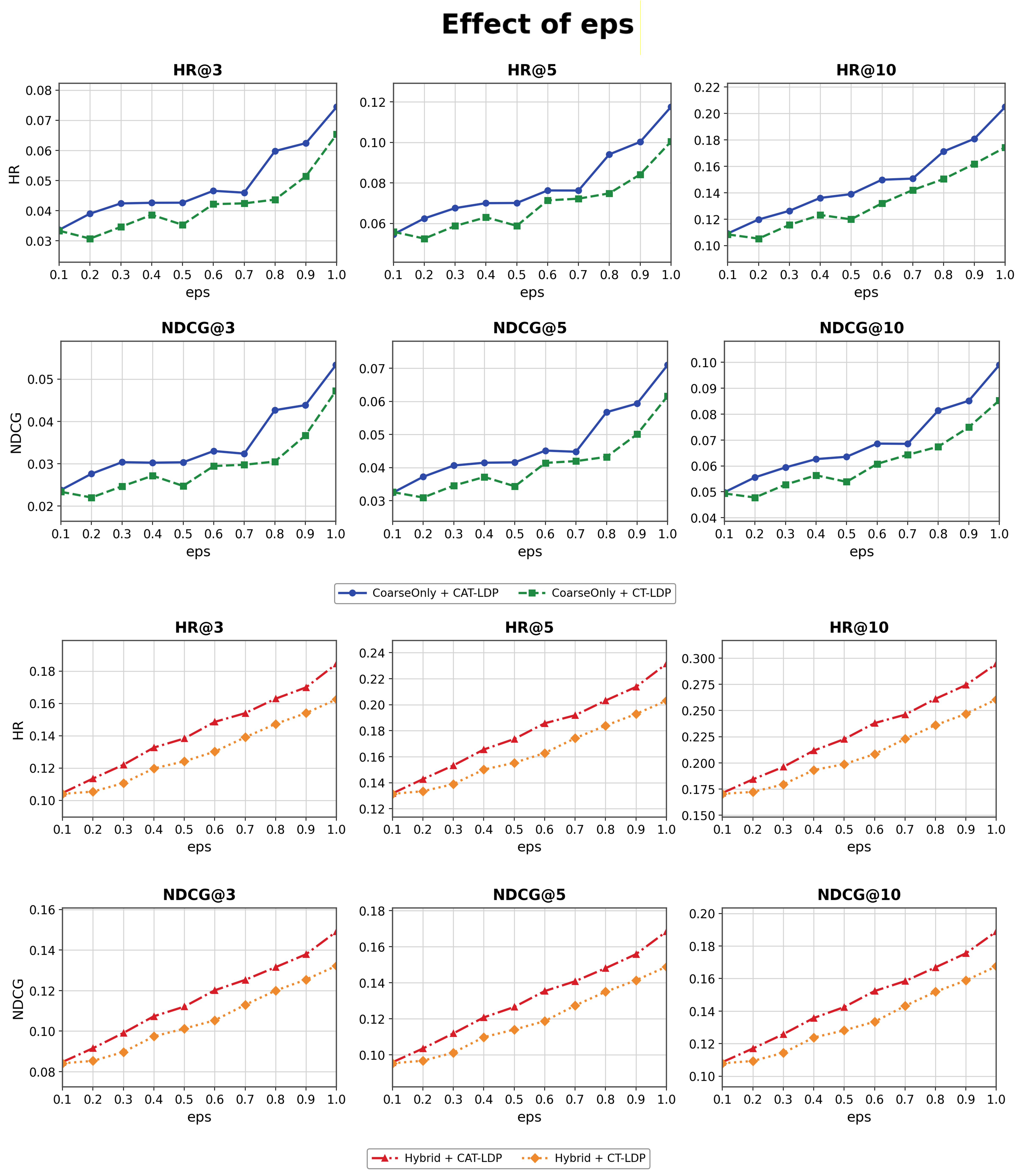}
    \caption{Top-K recommendation performance comparison on the Amazon Video Game 5-core dataset.}
    \label{fig:Adaptive_vs_Fixed}
\end{figure*}

To evaluate the effectiveness of the adaptive privacy budget allocation mechanism, this section conducts comparative experiments between CAT-LDP and its ablation variant, CT-LDP, which employs a fixed budget. The experiments investigate the performance trends of the models under both the CoarseOnly and HybridLocalRerank settings when the privacy budget $\epsilon$ varies in the range of 0.1 to 1.0. The selected evaluation metrics are HR@3, HR@5, HR@10, and NDCG@3, NDCG@5, NDCG@10. The resulting line charts are shown in Figure \ref{fig:Adaptive_vs_Fixed}.

From the trends in the line charts, the following observations can be made:

\begin{itemize}
    \item \textbf{Overall trend and budget sensitivity:} As the privacy budget $\epsilon$ increases (i.e., the injected randomized response noise decreases), the models under both settings exhibit a steady upward trend across all evaluation metrics. This aligns with the conventional trade-off between the degree of privacy protection and data utility in the differential privacy framework.
    
    \item \textbf{Mechanism gain in the coarse ranking stage:} Under the CoarseOnly setting, CAT-LDP with the adaptive budget outperforms CT-LDP with the fixed budget across all $\epsilon$ values and different truncation lengths $K$. This result indicates that by dynamically skewing the budget based on users' active features, the model can better preserve core preference signals under the same total privacy cost, thereby providing higher-quality feature inputs for clustering and coarse ranking on the cloud.
    
    \item \textbf{Advantage preservation in the reranking stage:} Under the Hybrid setting, which introduces local fine-grained reranking, CAT-LDP continues to consistently outperform CT-LDP across all tested metrics. This demonstrates that even when unperturbed true local historical data is introduced for Jaccard similarity computation during the reranking stage, the data quality improvement brought by the adaptive budget mechanism in the early profile construction stage can still effectively propagate and positively impact the final recommendation ranking list.
\end{itemize}

In summary, the adaptive privacy budget allocation mechanism demonstrates good robustness under our experimental settings, effectively improving the overall ranking performance of the framework under varying degrees of strict privacy constraints.

\section{Conclusion}

To address the high-dimensional sparsity and signal degradation issues in implicit feedback recommendation under Local Differential Privacy (LDP) constraints, this paper proposes a category tree-based cloud-local collaborative recommendation framework named CAT-LDP. Through an adaptive budget allocation mechanism, this framework skews the privacy budget toward the dimensions of users' active categories, improving the retention of effective signals in key feature dimensions under a fixed average budget constraint. By combining cloud-side clustering for coarse ranking and local metadata for reranking, it achieves a balance between privacy protection and ranking performance.

On the current Video Games dataset and under the $\epsilon=1.0$ setting, CAT-LDP outperforms both the fixed-budget ablation variant and the comparative baselines across all reported HR@K and NDCG@K metrics. The experimental results demonstrate that modeling in the category space and the cloud-local collaborative workflow help alleviate the noise amplification issue in long-tail sparse scenarios. Future work will further validate the generalizability of this framework across multiple datasets and under different evaluation protocols.

\balance
\bibliographystyle{IEEEtran}
\bibliography{references}

@article{koren2009matrix,
  author  = {Koren, Yehuda and Bell, Robert and Volinsky, Chris},
  title   = {Matrix factorization techniques for recommender systems},
  journal = {Computer},
  volume  = {42},
  number  = {8},
  pages   = {30--37},
  year    = {2009}
}

@inproceedings{salakhutdinov2007pmf,
  author    = {Salakhutdinov, Ruslan and Mnih, Andriy},
  title     = {Probabilistic matrix factorization},
  booktitle = {Advances in Neural Information Processing Systems},
  volume    = {20},
  pages     = {1257--1264},
  year      = {2007}
}

@inproceedings{rendle2009bpr,
  author    = {Rendle, Steffen and Freudenthaler, Christoph and Gantner, Zeno and Schmidt-Thieme, Lars},
  title     = {{BPR}: Bayesian personalized ranking from implicit feedback},
  booktitle = {Proceedings of the Twenty-Fifth Conference on Uncertainty in Artificial Intelligence ({UAI})},
  pages     = {452--461},
  year      = {2009},
  publisher = {AUAI Press}
}

@inproceedings{he2014practical,
  author    = {He, Xinran and Pan, Junfeng and Jin, Ou and Xu, Tianbing and Liu, Bo and Xu, Tao and Shi, Yanxin and Atallah, Antoine and Herbrich, Ralf and Bowers, Stuart and others},
  title     = {Practical lessons from predicting clicks on ads at {Facebook}},
  booktitle = {Proceedings of the Eighth International Workshop on Data Mining for Online Advertising},
  pages     = {1--9},
  year      = {2014}
}

@inproceedings{zhu2018learning,
  author    = {Zhu, Han and Li, Xiang and Zhang, Pengye and Li, Guozheng and Jie, Jianzhu and Chen, Xiaojiang and Gai, Kun},
  title     = {Learning tree-based deep model for recommender systems},
  booktitle = {Proceedings of the 24th {ACM} {SIGKDD} International Conference on Knowledge Discovery \& Data Mining},
  pages     = {1079--1088},
  year      = {2018}
}

@article{kasiviswanathan2011can,
  author  = {Kasiviswanathan, Shiva Prasad and Lee, Homin K. and Nissim, Kobbi and Raskhodnikova, Sofya and Smith, Adam},
  title   = {What can we learn privately?},
  journal = {SIAM Journal on Computing},
  volume  = {40},
  number  = {3},
  pages   = {793--826},
  year    = {2011}
}

@inproceedings{erlingsson2014rappor,
  author    = {Erlingsson, {\'U}lfar and Pihur, Vasyl and Korolova, Aleksandra},
  title     = {{RAPPOR}: Randomized aggregatable privacy-preserving ordinal response},
  booktitle = {Proceedings of the 2014 {ACM} {SIGSAC} Conference on Computer and Communications Security},
  pages     = {1054--1067},
  year      = {2014}
}

@article{shin2018privacy,
  author  = {Shin, Hyejin and Kim, Sungwook and Shin, Jun-Bum and Xiao, Xiaokui},
  title   = {Privacy enhanced matrix factorization for recommendation with local differential privacy},
  journal = {IEEE Transactions on Knowledge and Data Engineering},
  volume  = {30},
  number  = {9},
  pages   = {1770--1782},
  year    = {2018}
}

@article{li2025ldpmf,
  author  = {Li, Xiang and Zhou, Wang and Haq, Amin Ul and Khan, Shakir},
  title   = {{LDPMF}: Local differential privacy enhanced matrix factorization for advanced recommendation},
  journal = {Knowledge-Based Systems},
  volume  = {309},
  pages   = {112892},
  year    = {2025}
}

@inproceedings{gao2020dplcf,
  author    = {Gao, Chen and Huang, Chao and Lin, Dongsheng and Jin, Depeng and Li, Yong},
  title     = {{DPLCF}: Differentially private local collaborative filtering},
  booktitle = {Proceedings of the 43rd International {ACM} {SIGIR} Conference on Research and Development in Information Retrieval},
  pages     = {961--970},
  year      = {2020}
}

@article{bayesian_ldp_2025,
  title={Bayesian Local Differential Privacy for Implicit Feedback Recommendation},
  author={Tang, Hao and Wang, Yong and Li, Bo and Deng, Jiangzhou and Zhang, Zhiqiang},
  journal={Journal of Machine Learning and Information Security},
  volume={1},
  number={1},
  pages={6},
  year={2025}
}

@inproceedings{he2017neural,
  title={Neural Collaborative Filtering},
  author={He, Xiangnan and Liao, Lizi and Zhang, Hanwang and Nie, Liqiang and Hu, Xia and Chua, Tat-Seng},
  booktitle={Proceedings of the 26th International Conference on World Wide Web},
  series={WWW '17},
  pages={173--182},
  year={2017},
  publisher={International World Wide Web Conferences Steering Committee},
  address={Republic and Canton of Geneva, CHE},
  doi={10.1145/3038912.3052569}
}

@article{zhang2025privacy,
  title={Privacy-preserving recommendations with mixture model-based matrix factorization under local differential privacy},
  author={Zhang, Pengfei and Sun, Hong and Zhang, Zhikun and Cheng, Xiang and Zhu, Youwen and Zhang, Ji},
  journal={IEEE Transactions on Industrial Informatics},
  year={2025},
  publisher={IEEE}
}

@article{yang2024survey,
  title={Local differential privacy and its applications: A comprehensive survey},
  author={Yang, Mengmeng and Guo, Taolin and Zhu, Tianqing and Tjuawinata, Ivan and Zhao, Jun and Lam, Kwok-Yan},
  journal={Computer Standards \& Interfaces},
  volume={89},
  pages={103827},
  year={2024},
  publisher={Elsevier}
}

@inproceedings{hu2008implicit,
  author    = {Hu, Yifan and Koren, Yehuda and Volinsky, Chris},
  title     = {Collaborative Filtering for Implicit Feedback Datasets},
  booktitle = {Proceedings of the 2008 Eighth IEEE International Conference on Data Mining},
  pages     = {263--272},
  year      = {2008},
  doi       = {10.1109/ICDM.2008.22}
}

@inproceedings{calandrino2011privacy,
  author    = {Calandrino, Joseph A. and Kilzer, Ann and Narayanan, Arvind and Felten, Edward W. and Shmatikov, Vitaly},
  title     = {“You Might Also Like:” Privacy Risks of Collaborative Filtering},
  booktitle = {2011 IEEE Symposium on Security and Privacy},
  pages     = {231--246},
  year      = {2011},
  doi       = {10.1109/SP.2011.40}
}

@article{xin2023behavior,
  author     = {Xin, Xin and Yang, Jiyuan and Wang, Hanbing and Ma, Jun and Ren, Pengjie and Luo, Hengliang and Shi, Xinlei and Chen, Zhumin and Ren, Zhaochun},
  title      = {On the User Behavior Leakage from Recommender System Exposure},
  journal    = {ACM Transactions on Information Systems},
  volume     = {41},
  number     = {3},
  articleno  = {57},
  pages      = {1--25},
  year       = {2023},
  doi        = {10.1145/3568954}
}

@article{mullner2023dpreview,
  author  = {M{\"u}llner, Peter and Lex, Elisabeth and Schedl, Markus and Kowald, Dominik},
  title   = {Differential Privacy in Collaborative Filtering Recommender Systems: A Review},
  journal = {Frontiers in Big Data},
  volume  = {6},
  pages   = {1249997},
  year    = {2023},
  doi     = {10.3389/fdata.2023.1249997}
}

@article{ali2025privacysurvey,
  author  = {Ali, Waqar and Zhou, Xiangmin and Shao, Jie},
  title   = {Privacy-preserved and Responsible Recommenders: From Conventional Defense to Federated Learning and Blockchain},
  journal = {ACM Computing Surveys},
  volume  = {57},
  number  = {5},
  page = {1--35},
  year    = {2025},
  doi     = {10.1145/3708982}
}

@article{zhang2024attribute,
  author  = {Zhang, Shijie and Yuan, Wei and Yin, Hongzhi},
  title   = {Comprehensive Privacy Analysis on Federated Recommender System Against Attribute Inference Attacks},
  journal = {IEEE Transactions on Knowledge and Data Engineering},
  volume  = {36},
  number  = {3},
  page    = {987--999},
  year    = {2024},
  doi     = {10.1109/TKDE.2023.3295601}
}

@incollection{mullner2024impact,
  author    = {M{\"u}llner, Peter and Lex, Elisabeth and Schedl, Markus and Kowald, Dominik},
  title     = {The Impact of Differential Privacy on Recommendation Accuracy and Popularity Bias},
  booktitle = {Advances in Information Retrieval},
  series    = {Lecture Notes in Computer Science},
  volume    = {14611},
  pages     = {466--482},
  year      = {2024},
  publisher = {Springer},
  doi       = {10.1007/978-3-031-56066-8_33}
}

@inproceedings{ni2019just,
  title={Justifying recommendations using distantly-labeled reviews and fine-grained aspects},
  author={Ni, Jianmo and Li, Jiacheng and McAuley, Julian},
  booktitle={Proceedings of the 2019 conference on empirical methods in natural language processing and the 9th international joint conference on natural language processing (EMNLP-IJCNLP)},
  pages={188--197},
  year={2019}
}

\end{document}